\pdfoutput=1

\documentclass[letterpaper, 10 pt, conference]{ieeeconf}  

\IEEEoverridecommandlockouts                              
\usepackage[utf8]{inputenc}
\usepackage[T1]{fontenc}
\usepackage{graphicx}
\usepackage{textgreek}
\usepackage{cite}
\usepackage{multirow}
\usepackage[colorlinks  = true,
            citecolor   = blue,
            linkcolor   = blue,    
            urlcolor    = blue,   
            breaklinks  = true,    
            pdfauthor   = {Will\ Coon},
            pdftitle    = {ezsleep},
            pdfsubject  = { },
            pdfkeywords = { },
            plainpages  = false,
            pdfpagelabels
           ]{hyperref}
\usepackage{dsfont}

\begin{document}

\title{\LARGE \bf
\texttt{StARS DCM}: A Sleep Stage-Decoding Forehead EEG Patch for Real-time Modulation of Sleep Physiology
}



\author{William G. Coon$^{1,}$$^{2}$, Preston Peranich$^{1}$, and Griffin Milsap$^{1,}$$^{2}$
\thanks{*This work was supported by the John Hopkins University Applied Physics Laboratory. Corresponding authors: \texttt{will.coon@jhuapl.edu} and \texttt{griffin.milsap@jhuapl.edu}}
\thanks{$^{1}$Johns Hopkins University Applied Physics Laboratory,
        11100 Johns Hopkins Rd., Laurel, MD 20723, USA
        }%
\thanks{$^{2}$Johns Hopkins University Whiting School of Engineering,
        11100 Johns Hopkins Rd., Laurel, MD 20723, USA}
}

\maketitle
\thispagestyle{empty}
\pagestyle{empty}

\begin{abstract}
The System to Augment Restorative Sleep (StARS) is a modular hardware/software platform designed for real-time sleep monitoring and intervention. Utilizing the compact DCM biosignal device, StARS captures electrophysiological signals (EEG, EMG, EOG) and synchronizes sensor data using the \texttt{ezmsg} real-time software framework. StARS supports interventions such as closed-loop auditory stimulation and dynamic thermal modulation guided by sleep-stage decoding via advanced neural network models and transfer learning. Configurable with a lightweight EEG forehead patch or wearable sensors like smart rings, StARS offers flexible, low-burden solutions for EEG, BCI, and sleep-enhancement research and applications. The open-source DCM patch further enables customizable EEG device development.

\end{abstract}

%

\begin{figure*}[thpb]
      \centering
      \includegraphics[width=1\textwidth]{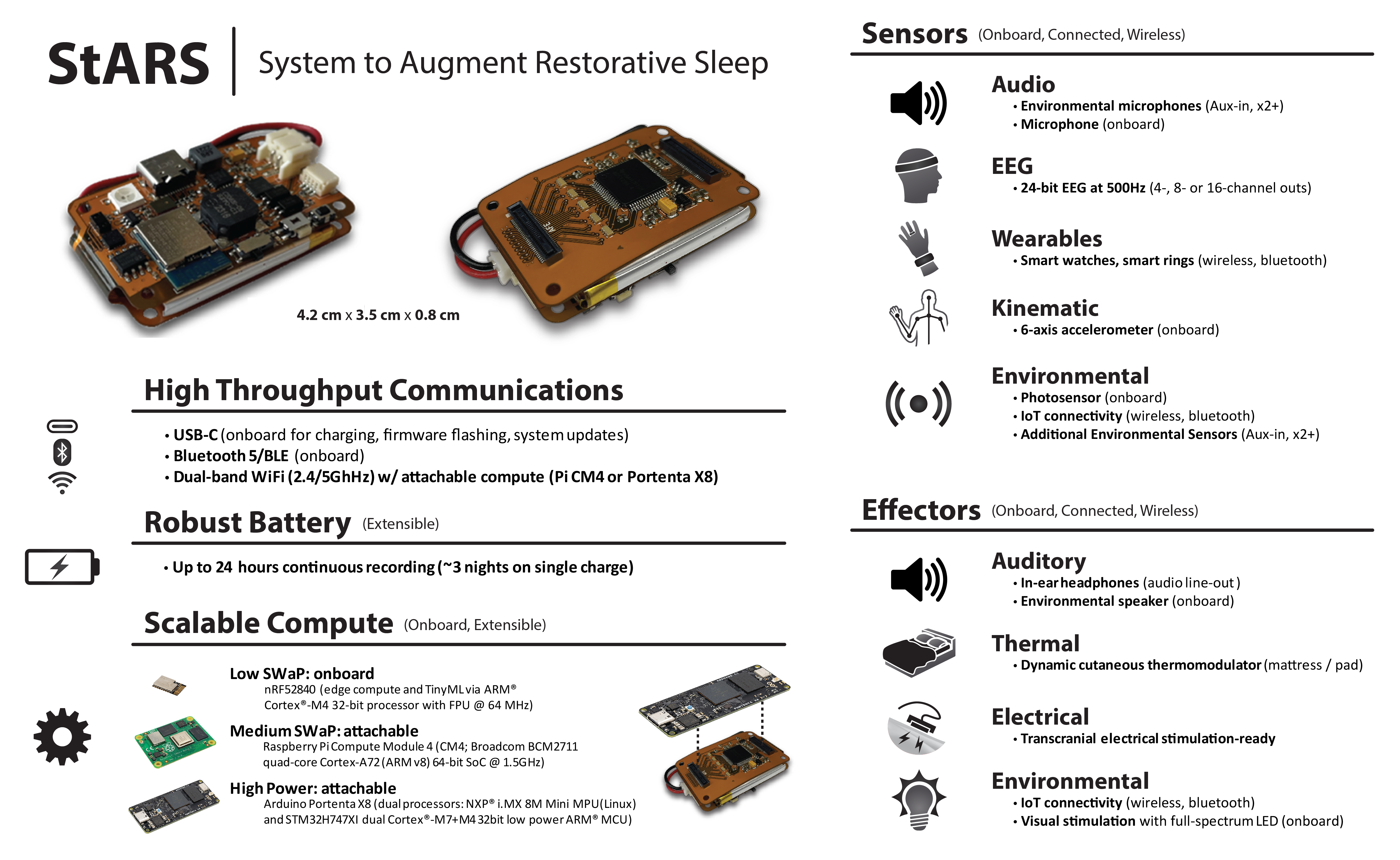}
      \caption{\textbf{StARS} can utilize the DCM (JHU/APL), an open-source forehead EEG patch designed for realtime processing and sleep research.}
      \label{fig:stars}
    \end{figure*}

\section{OVERVIEW}
\label{intro}

Sleep is fundamental to good health and optimal performance, yet Americans are chronically sleep-deprived\cite{krueger2009sleep} and America’s armed services even more so \cite{mysliwiec2013sleep, luxton2011prevalence}. Inadequate sleep impairs cognitive function \cite{killgore2010effects}, learning capacity \cite{curcio2006sleep}, memory \cite{newbury2021sleep,diekelmann2010memory}, immune system function \cite{bryant2004immune}, neural waste metabolite clearance \cite{xie2013sleep}, physical recovery \cite{rae2017physrecovery}, cardiovascular health \cite{mullington2009cardiovascular}, and growth and metabolism \cite{knutson2007metabolic}. It has been linked to cognitive decline in old age \cite{scullin2015healthyaging}, specific maladies of the brain like Alzheimer’s Disease and dementia\cite{spira2014cogdeclineaging}, and may even underlie the increased risk of these neurodegenerative sequelae in those who have suffered a traumatic brain injury (TBI) \cite{piantino2022glymphneuroaging,peters2023glymphatic,hablitz2021glymphatic}. While many of the functions above have been linked to non-rapid eye movement (NREM) sleep, rapid eye movement (REM) sleep plays a critical role in processing emotional memory \cite{vanderhelm2011emotion,wassing2019restless}, and the dysfunction of this process may contribute to post-traumatic stress disorder (PTSD) \cite{rho2023emotional,mellman2002rem}.  Systems that can initiate and maintain different sleep states on-demand offer whole new avenues to treat disease \cite{lubin1974recuperative,geiser2020targeting}, and may even be able to increase sleep’s efficiency, achieving its restorative effects in less time.

To these ends, JHU/APL has developed a modular hardware/software ecosystem, configured for these purposes as JHU/APL’s System to Augment Restorative Sleep (StARS), that can support standardization and research in sleep and accelerate the development of sleep-enhancing technologies.  StARS is a system of modular attachables (sensors, compute, effectors) that can accurately decode sleep in real time and support a wide variety of sleep stimulation protocols.  Its software backbone has been built using a publicly available, JHU/APL built, high-throughput messaging framework called \texttt{ezmsg} (pip-installable or available for free at github.com/ezmsg-org/ezmsg). \texttt{ezmsg} employs a publisher-subscriber (“pub/sub”) model, similar to and cross-compatible with Robot Operating System (ROS) and Lab Streaming Layer (LSL), that is optimized for realtime applications and the coordination of multiple systems (ex. sensors/effectors/compute), but with lighter computational costs and faster and more efficient messaging than popular alternatives.  

StARS is an adaptable technology set that can integrate multiple EEG base platforms including the DCM, an APL-designed and manufactured miniaturized sensor suite, approximately the size of two US quarters (Fig. \ref{fig:stars}D).  The DCM in its simplest configuration can collect up to 16 channels of electrophysiological input such as electroencephalography (EEG), electromyography (EMG), electro-oculography (EOG), etc., at 24-bit resolution.  Its snap-based electrode connector allows for auxiliary inputs, wired electrode extensions, and modular sensor boards that can be configured for any desired sensor layout and affixed to the wearer with industry standard EKG-style conductive gel electrode stickers.  The DCM additionally contains an IMU (motion sensor), microphone, ambient light/IR sensor, haptic driver and LEDs.  StARS’ modularity allows a user to control a variety of effectors to modulate the brain activity and physiological state of the wearer based on closed-loop sensing of brain state (ex. sleep stage), neurophysiology (ex. EEG slow wave detection), behavior (ex. movement, vocalization), or environmental variables (ex. temperature, noise, light).  

StARS effectors designed to improve sleep quality include closed-loop auditory stimulation, and bedding temperature modulation conditioned on the wearer’s realtime-decoded ongoing sleep stage dynamics.  StARS is also being tested using automated neural network-based sleep stage decoding applied to one of two signal sources: EEG signals, or signals available to peripheral wearables like smart rings and smart watches, such as heart rate, heart rate variability (HRV), motion, and breathing.  While non-EEG, peripheral/wearable sleep stage decoders are generally accepted to be far less accurate, JHU/APL has developed novel self-supervised and transfer learning approaches, adapted from the field of audio speech processing, that significantly enhance the accuracy of these decoders \cite{coon2025transfer}.  This fundamental advance makes possible a compact, low size, weight, and power (SWaP) rendition of StARS consisting only of a smart ring, a smartphone app, and thermoregulating bedding or garments.

\begin{figure*}[thpb]
      \centering
      \includegraphics[width=0.99\textwidth]{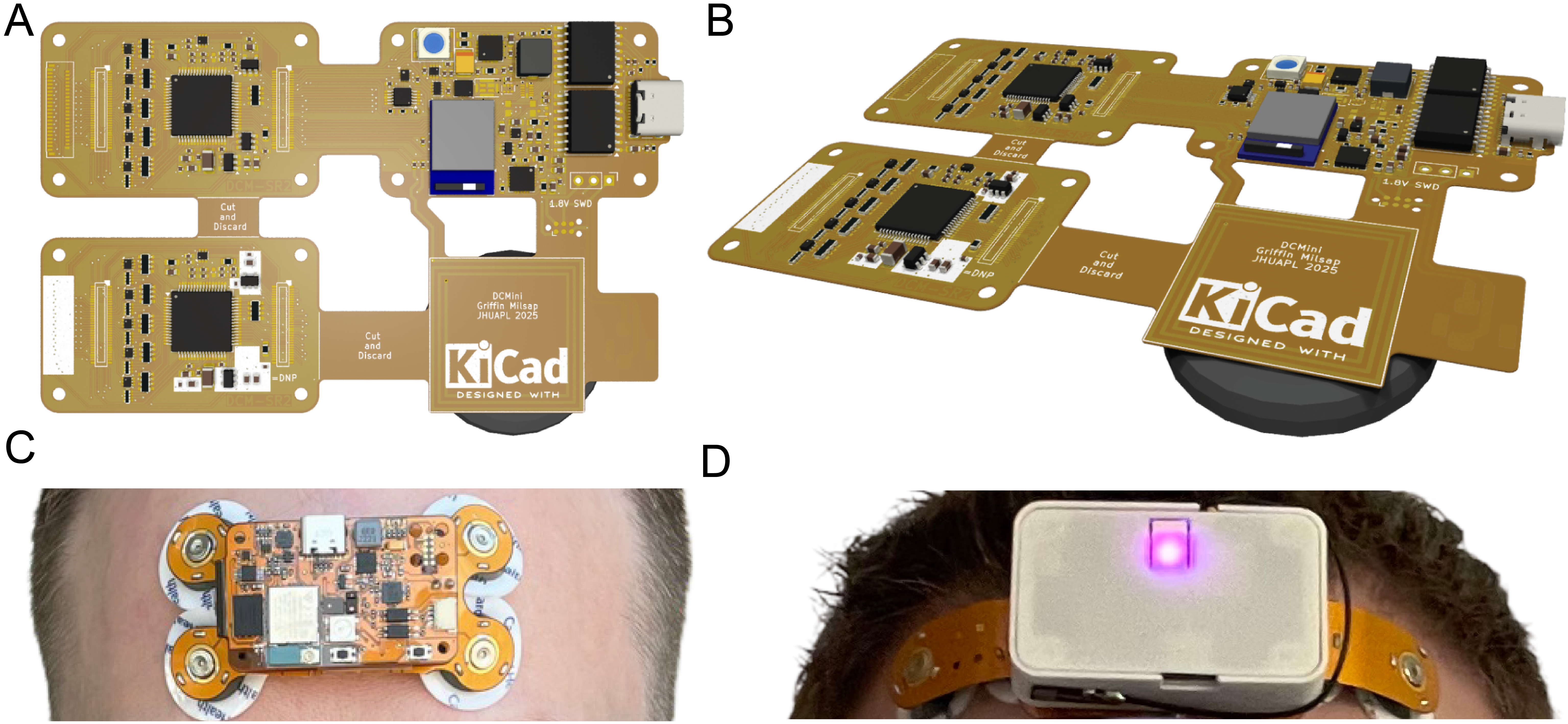}
      \caption{\textbf{Flexible configurations.} The DCM forehead patch can be easily reconfigured to adopt one of many possible hardware configurations. \textbf{A.} the DCM board is manufactured on flexible PCB substrates that allow the device's board modules to fold over one another or to lie flat, allowing the user to achieve a compact stack of boards (as in C), or a flat conformable form factor that adheres to the contours of the forehead. Additionally, optional functionality like NEC or daisy-chained ADC boards can be easily removed if not needed for the intended application. \textbf{B.} DCM's state-of-the-art dense storage coin cell battery, seen prominently on the lower right of the image as a dark gray disc, allows for full-length recordings with minimal mass/volume add from the battery. \textbf{C.} Configurable and attachable electrode boards (not shown in A or B) allow multiple electrode sensor montages and adaptable numbers of sensor contacts. Differential referencing can also be set at the hardware level for grounded, referenced, and/or bipolar electrode channels derivations. \textbf{D.} DCM configured with a five-sensor electrode board, configured for sleep EEG recordings.  Pressing the DCM's single font-facing button turns on the device, and a double-tap initiates a recording stored directly to its onboard microSD card. }
      \label{fig:dcmHW}
    \end{figure*}

\begin{figure*}
    \centering
    \includegraphics[width=1\textwidth]{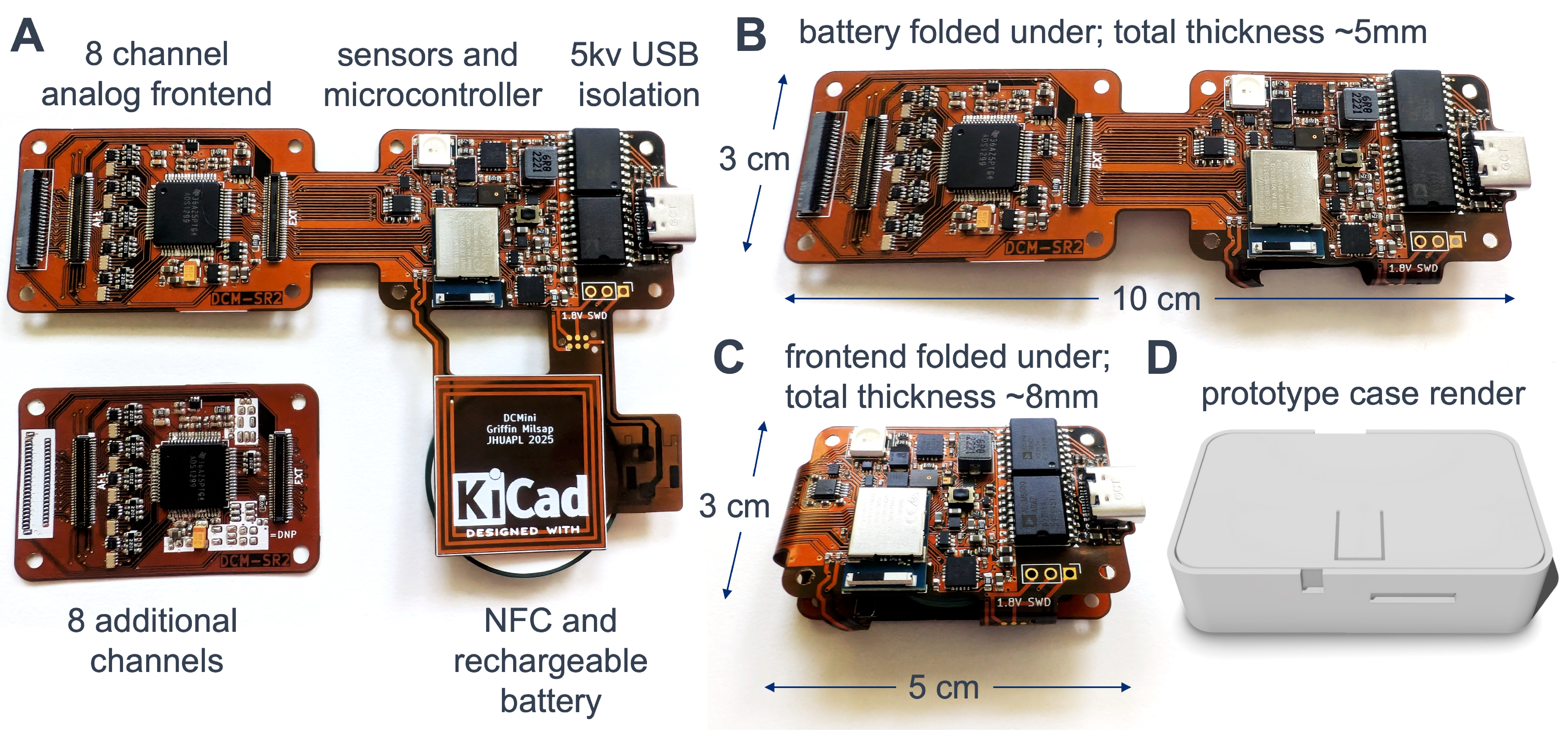}
    \caption{\textbf{Modular design, configurable form factor}: Functional prototype of the DCM. \textbf{A.} PCB has been fabricated, populated, and parted out. Capabilities and sections of the board are labeled. \textbf{B.} Additional analog front end has been attached below main analog frontend for 16-channel recording capability; battery/NFC antenna has been folded under for a thinner form factor with a 10x3 cm footprint, suitable for placement on the forehead. \textbf{C.} Analog frontend can be further folded under the digital side of the board resulting in an even smaller footprint at the expense of another 3 mm of device thickness. Electrode daughterboards connect to the analog frontend either via the board-to-board mezzanine connectors on the analog frontend, or via the FPC connector on the far left of the analog frontend as pictured in configurations A and B. \textbf{D.} A prototype case render for the fully folded version of the DCM.}
    \label{fig:hiresphotos}
\end{figure*}


\section{An open-source forehead EEG patch: the DCM}

The DCM is a forehead patch-style, micro-EEG unit that facilitates synchronized multi-sensor data logging and real-time data streaming. The primary sensor onboard is the ADS1299 precision instrumentation amplifier, capable of capturing 4 to 16 channels of high-resolution electrophysiological signals, currently optimized for neural data acquisition via scalp EEG. The device also includes a 6-axis inertial measurement unit (IMU) for precise motion tracking; a microphone designed to capture speech, behaviorally relevant audible physiology such as respiratory activity and snoring, and ambient environmental audio; and an ambient light sensor to monitor environmental conditions.

Data storage and logging are managed through an integrated microSD card, ensuring robust and reliable data capture. The DCM provides intuitive user feedback through a multi-color LED indicator (Fig. \ref{fig:dcmHW}D) and haptic feedback via an embedded linear resonant actuator (LRA, vibrator). To expand functionality, the DCM supports daughterboards for additional sensor integration. Safety during wired operation and battery charging is assured through a built-in 5kV USB isolation barrier, enabling safe use even while electrically connected to the wearer.

For seamless device pairing and enhanced usability, the DCM employs near-field communication (NFC) technology, allowing users to quickly connect the device to smartphones or other compatible electronics via a simple tap gesture, leveraging Bluetooth connectivity.
DCM’s design emphasizes flexibility and ergonomics, featuring a slim, conformable form-factor achieved through a carefully optimized flexible PCB substrate. The device can comfortably conform to the forehead or fold into a compact footprint as needed. The core processing component is an ultra-low-power, multi-protocol wireless-enabled microcontroller (nRF52840, Nordic Semiconductors), supporting approximately 5 days of continuous multi-sensor logging with intermittent EEG recordings, or more than 8-10+ hours of continuous EEG recording per charge. A compact 300 mAh coin-cell LiPo battery powers the device, capable of rapid charging from empty to full in approximately 20 minutes.

DCM is also economically optimized, with a total bill of materials cost of roughly \$180 USD per unit, with substantial cost reductions achievable at scale. The PCB layout has been meticulously refined to require only a standard 6 mil trace/6 mil spacing, 2-layer flexible PCB fabrication process, compatible with affordable prototyping services such as OSHPark.

We are committed to fostering open scientific innovation, and the DCM platform will soon be released as open hardware. Board design files created using free and open-source CAD software (KiCAD), along with firmware under permissive licensing, will be freely available via GitHub. Project files will also be accessible through commercial assembly services such as Tindie. Ultimately, DCM aims to democratize access to synchronized multi-modal data acquisition, seamlessly integrating EEG, EMG, ECG, and other biopotential measurements into unified data streams.

\section{Increasing Slow Waves / Slow Wave Sleep with STARS}
\label{sws}

Much of sleep’s restorative nature has been linked to Slow Wave Sleep (SWS), a stage of NREM sleep characterized by large-amplitude low-frequency oscillations observable in the scalp electroencephalogram (EEG) called ‘Slow-Waves’. Slow waves largely occur during the first two sleep cycles and SWS is accompanied by sharp declines in core body temperature after falling asleep \cite{campbell1994rapid,raymann2008skin}. Growth hormone is also excreted during SWS \cite{born1988significance} and reduced SWS is correlated with complaints of non-restorative, shallow sleep \cite{dijk1990electroencephalogram,crenshaw1999slow}. It is also widely accepted that SWS plays a role in immune function, memory consolidation, age-related cognitive decline and dementia \cite{irwin2019implications,himali2023association,girardeau2021brain,nedergaard2020glymphatic}. This may be due in part to the recent discovery that SWS drives the glymphatic system (GS), which facilitates the exchange of cerebrospinal and interstitial fluids in the brain \cite{nedergaard2020glymphatic}. The GS clears metabolic waste from the brain (e.g., amyloid-β) and may regulate neuroinflammation \cite{nedergaard2020glymphatic,chong2022sleep}. All these reasons make slow waves and SWS a compelling target for optimizing sleep quality.


\subsection{Acoustic slow-wave stimulation (aSTIM) }

aSTIM is an emerging, non-invasive and scalable technology that boosts the amplitude of slow waves (SWs) and overall Slow Wave Activity (SWA: total power in the 0.4-5hz EEG range) by delivering precisely timed brief (50ms), ‘pink noise’ acoustic stimuli \cite{ngo2013auditory}. Two recent meta-analyses of 16 studies concluded that aSTIM improved memory consolidation and cognitive performance in adults \cite{stanyer2022impact,wunderlin2021auditorymemory}. Moreover, promising work indicates that aSTIM may modulate other SWS-regulated processes, including immune38 and autonomic functions \cite{grimaldi2019strengthening}.  In typical aSTIM protocols, a 50ms burst of ‘pink-noise’ is applied during the UP portion of a slow wave (400ms after threshold crossing), which increases the peak amplitude and area under the curve of the wave.  StARS uses a similar protocol.  

\subsection{Enhancing slow waves with dynamic bedding temperature yoked to sleep cycle patterns}

Passively timed body cooling is one of the best-studied methods of increasing SWS and SW-density. Sleep onset coincides with the maximal decline in core body temperature (CBT), initiated via heat-loss through the skin \cite{krauchi2007thermophysiological}. Meta-analyses demonstrate that elevating CBT ~1.64-2.58°C, 2 to 2.5 hours prior to habitual bedtime (achieved by 30 min+ hot baths and/or intense exercise) results in increased SWS and SWA \cite{raymann2008skin,naylor2000daily,horne2013exercise,tai2021hot,dorsey1996effects,bunnell1988passive}. The presumed mechanism is that the spike in CBT augments the normal CBT drop that occurs at sleep onset, which in turn increases SW-production \cite{bunnell1988passive}. One study suggested that sleeping on a heat-conducting mattress, which lowers CBT by conducting heat from the body, significantly increased SWS by 16\% \cite{krauchi2007thermophysiological,krauchi2018sleep}. This approach has the advantage of requiring limited behavioral change, but lacks the ability to control and optimize the cooling process.  Recent technology has developed water-cooled conductive materials, which can control the precise temperature i.e., Active Body Cooling (ABC) of the cooling surface, but they have not previously been integrated with a noninvasive, forehead sticker-encapsulated brain-computer interface to optimally time cooling according to sleep physiology.  To our knowledge, StARS is the first system to bridge this gap. 


Thermal modulation protocols aim to guide the body’s natural change in core body temperature when transitioning from wakefulness to deep, slow wave sleep. StARS employs a water-filled mattress pad to quickly change the temperature of the sleeper’s bedding, changing from a neutral temperature at bedtime start, to a cooler temperature at desired times relative to decoded sleep stage dynamics.  Traditionally this is done by manually pushing a button to mark bedtime, after which the mattress will initiate cooling at a fixed time delay (ex. 20 minutes after bedtime).  Since this does not take into account the sleeper’s state (they may still be awake, for example), there is no guarantee that the timing will work out as intended.  Yoking the cooling to our realtime sleep stage decoder provides a guarantee that cooling is initiated and dynamically modulated at the appropriate times, regardless of variability in sleep onset latency.  

\section{Summary}

The StARS system is just one example of the types of sophisticated applications the DCM can support and DCM has already been validated for use in automated sleep stage classifiers \cite{coon2025ezscore}.  Similarly, acoustic stimulation and environmental control based on accurate real-time knowledge of the sleeper's ongoing sleep stage dynamics are but two of many possible effector-driven interventions that could be included in the StARS system itself to optimize sleep. With the release of the open-source DCM, we hope to empower the research community with this same type of flexible utility and rapid-prototyping potential to develop the next generation of non-invasive neural interface applications.

\bibliographystyle{IEEEtran}
\bibliography{sleep}

\end{document}